\documentclass{pasj00}

\begin{document}
\SetRunningHead{Egusa, Sofue, and Nakanishi}
{Offsets between H$\alpha$ and CO: Pattern Speed Determination}
\Received{2004/09/08}
\Accepted{2004/10/09}

\title{Offsets between H$\alpha$ and CO arms of a spiral galaxy 
NGC 4254: A New Method for Determining
       the Pattern Speed of Spiral Galaxies}

\author{Fumi \textsc{Egusa}, Yoshiaki \textsc{Sofue}, 
and Hiroyuki {\sc Nakanishi}}
\affil{Institute of Astronomy, University of Tokyo}
\email{fegusa@ioa.s.u-tokyo.ac.jp, 
sofue@ioa.s.u-tokyo.ac.jp, hnakanis@ioa.s.u-tokyo.ac.jp}

%

\KeyWords{galaxies: fundamental
parameters---galaxies: individual(NGC 4254, 
M99)---galaxies: spiral---ISM: molecules---ISM: H{\sc ii} 
region} 

\maketitle

\begin{abstract}
 We examined offsets between H{\sc ii} regions and 
molecular clouds belonging to spiral arms of 
a late type spiral galaxy NGC 4254 (M99).
 We used a high resolution $^{12}$CO ($J$=1-0) image obtained 
by Nobeyama Millimeter Array (NMA) 
and an H$\alpha$ image. 
 We derived angular offsets ($\theta$) in the galactic disk, and 
found that these offsets show a linear dependence on  
the angular rotation velocity of gas ($\Omega_{\rm G}$). 
 This linear relation can be expressed by an equation: 
$\theta =(\Omega_{\rm G}-\Omega_{\rm P})\cdot t_{\rm H\alpha}$, 
where $\Omega_{\rm P}$ and $t_{\rm H\alpha}$ are constant. 
 Here, $\Omega_{\rm P}$ corresponds to the pattern speed of 
spiral arms and $t_{\rm H\alpha}$ is interpreted as the timescale 
between the peak compression of the molecular gas in spiral arms 
and the peak of massive star formation.
 We may thus determine $\Omega_{\rm P}$ and $t_{\rm H\alpha}$ 
simultaneously by fitting a line to our $\theta - \Omega_{\rm G}$ plot, 
if we assumed they are constant.
 From our plot, we obtained 
$t_{\rm H\alpha} = ( 4.8 \pm 1.2 )\times 10^6$ yr 
and $\Omega_{\rm P} = 26^{+10}_{-6} ~{\rm km~s^{-1}~kpc^{-1}}$, 
which are consistent with previous studies.
 We suggest that this $\theta - \Omega_{\rm G}$ plot can be a 
new tool to determine the pattern speed 
and the typical timescale needed for star formations.
\end{abstract}

\section{Introduction}
\subsection{Pattern Speed}
 Pattern speed ($\Omega_{\rm P}$) is defined as 
an angular rotation velocity of spiral pattern.
 According to the spiral density wave theory \citep{LS64}, 
$\Omega_{\rm P}$ determines the location and even the existence of 
resonances, which would greatly affect the morphology 
and kinematics of a galaxy.
 The results of numerical simulations also show 
the large dependence of spiral structure on 
$\Omega_{\rm P}$ \citep{Semp95}. 
 Despite its importance on the study of galaxies, 
it cannot be determined directly from observations, 
since the pattern structure is not a material 
but a density wave propagating through the disk. 

 Several methods have been proposed to date for 
the determination of pattern speed.
 \citet{TW84} presented the method 
using the continuity equation for the surface brightness 
of galaxies.
 \citet{Canz93} showed that the residual velocity field, 
obtained by subtracting the axisymmetric component from 
observed velocity field, has a difference 
in pattern between the inside and outside of the corotation.
 \citet{CB90} pointed out that where the arm to interarm ratio of 
star formation efficiency has a sharp dip could be thought as 
the corotation.
 These methods work well for some galaxies, 
but there are still inconsistencies or large uncertainties. 

\subsection{CO and H$\alpha$}
 In this letter,
we examined offsets between H$\alpha$ and CO arms of a 
spiral galaxy. 
 We suggest that we can use them as a new method for the 
determination of pattern speed and 
typical timescale for star formation.

 The H$\alpha$ emission is the tracer of H{\sc ii} region, 
which surrounds newly born massive stars, 
while the CO emission is the tracer of molecular gas.
 According to the galactic shock wave theory 
\citep{Fuji68, Robe69},
most molecular clouds are formed and then condensed 
by the compression owing to the spiral shock wave 
and thus belong to the spiral arm.
 Since galactic scale star formations take place 
in these molecular arms, 
some molecular gases are used to form stars, 
and the others are 
dissociated by UV emission from massive stars 
born in that star formation.
 In consequence, molecular clouds disappear soon after their formation 
and their position appears to be fixed to the spiral pattern, 
though individual molecular clouds move at the speed of gas.
 Therefore, the position of spiral pattern and H$\alpha$ arm
can be measured by using that of 
a molecular arm and an ensemble of H{\sc ii} regions respectively.
 In other words, offset between the arms of H$\alpha$ and CO represents 
the difference between the speed of gas and spiral pattern, 
as well as the time needed for star formation from molecular clouds.

 This offset has been found in many spiral galaxies 
(eg. \cite{RK90}).
 We can easily see it in a $B$-band image 
as a displacement of an arm's peak and a dust lane, 
since the light of $B$-band is mainly emitted by the young OB star 
and the distribution of dust lane and molecular gas are almost the same.
 However, this offset has not been studied quantitatively yet, 
since the resolution of the CO image was too large to 
examine detail structures of spiral arms.

\section{A Method for Determining the Pattern Speed}
 We assume that 
the pattern of spiral is rigid, and that 
the gas rotates in circular orbit. 
 Then we define a timescale $t_{\rm H\alpha}$ 
as an {\it average} time for H$\alpha$ arm 
to develop from a galactic-shock-compressed molecular arm. 
 If the physical processes of star formation 
are not extremely different in the spiral disk, 
this timescale can be regarded as a constant parameter 
representing a typical value of the entire disk.

\begin{figure}
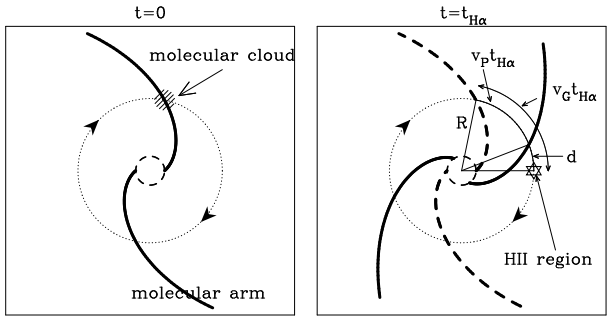

  \begin{center}
    \FigureFile(80mm,80mm){draw.ps}
  \end{center}
  \caption{The basic idea of our method. 
 If we observe a face-on spiral galaxy at $t=0$ (the left panel), 
the same galaxy will be observed  
as the right panel at $t=t_{\rm H\alpha}$.
 The thick solid lines are molecular arms at the time t of each panel.
 The thick dashed lines in the right panel (t=$t_{\rm H\alpha}$) 
show the position of molecular arms in the left panel (t=0).
 The offset distance between H$\alpha$ and CO arm is $d$, 
expressed in equation (\ref{eq.d}).}
\label{fig.draw}
\end{figure}

 Figure \ref{fig.draw} illustrates this idea  
for the inside of the CR.
 If we observe a face-on spiral galaxy at $t=0$ (the left panel), 
at $t=t_{\rm H\alpha}$ the same galaxy will be observed  
as the right panel 
and the offset distance between the arms of H$\alpha$ and CO, 
$d$, can be written as
\begin{equation}
 d=\Big(\frac{v_{\rm G}}{\rm km~s^{-1}}\Big)
  \Big(\frac{t_{\rm H\alpha}}{\rm s}\Big)
  -\Big(\frac{v_{\rm P}}{\rm km~s^{-1}}\Big)
  \Big(\frac{t_{\rm H\alpha}}{{\rm s}}\Big)
  \quad{\rm [km]},
\label{eq.d}
\end{equation}
where $v_{\rm G}$ is the velocity of gas, 
and $v_{\rm P}$ is the velocity of pattern.
 We adopt this expression, in which $d$ becomes a positive value, 
since most of molecules exist in a rather central part of 
the disk and thus they seem to be in the inside of the CR. 
 If molecular gas is more extended and 
the outside of the CR should be taken into account, 
H$\alpha$ arm will be seen on the concave side of CO
arm and $d$ will be negative.

 Dividing both sides of equation (\ref{eq.d}) by radius $R$ [kpc],  
we obtain
\begin{eqnarray}
\theta = \Big\{ 
   \Big(\frac{\Omega_{\rm G}}{\rm km~s^{-1}~kpc^{-1}}\Big)
   -\Big(\frac{\Omega_{\rm P}}{\rm km~s^{-1}~kpc^{-1}}\Big)
\Big\}
\nonumber\\
   \times\Big(\frac{t_{\rm H\alpha}}{\rm s}\Big)
   \quad{\rm [km~kpc^{-1}]},
\label{eq.ofst}
\end{eqnarray}
where $\Omega_{\rm G}\equiv v_{\rm G}/R$, 
$\Omega_{\rm P}\equiv v_{\rm P}/R$, and
$\theta$ is the azimuthal offset.
 This equation shows the relation  
between two observables, $\Omega_{\rm G}$ and $\theta$, 
and we can rewrite it as
\begin{equation}
 \theta \simeq 0.58 (\Omega_{\rm G}-\Omega_{\rm P})~t_{\rm H\alpha} 
\label{eq.linear}
\end{equation}
where $\theta$ is in degree and $t_{\rm H\alpha}$ is in $10^7$ yr.
 If we assume that $t_{\rm H\alpha}$ is constant in a certain range of 
radius in a galaxy,
$\theta$ becomes a linear function of $\Omega_{\rm G}$,
since we assume the rigid pattern, in other words, 
constant $\Omega_{\rm P}$. 
 Therefore, by plotting $\theta$ against $\Omega_{\rm G}$
and fitting them with a line, both $\Omega_{\rm P}$ and 
$t_{\rm H\alpha}$ can be determined at the same time.
 In addition, the linearity of this plot can verify the validity 
of the assumptions we made.
 We discuss this validity in section 4.

\section{Data}
 We observed a SAc galaxy NGC 4254 in the $^{12}$CO ($J$=1-0) line 
using Nobeyama Millimeter Array 
during a long-term project, ``Virgo high resolution CO survey'' 
\citep{ViCS1}.
 The spatial resolution is about 2 arcsec, 
which corresponds to $\sim 160$ pc at a distance 
from the Virgo cluster of 16.1 Mpc.
 This resolution is small enough to trace the spiral arms 
with typical width of 1 kpc.
 For more detailed information about this galaxy 
and the CO observation, see \citet{ViCS3}.

 \citet{Koop01} presented broadband {\it R} 
and narrowband H$\alpha$ images of 
63 spiral galaxies in the Virgo cluster.
 The resolution of the image is 1.2 arcsec, 
comparable to that of our CO data.
 We determined the coordinate of the H$\alpha$ image 
using the Galactic stars seen in the 
{\it R} band image, 
since these two images had the same field of view. 
 The typical uncertainty of this coordinate fitting is about 1 arcsec, 
smaller than the resolution of the images.
\begin{figure}
  \begin{center}
    \FigureFile(80mm,80mm){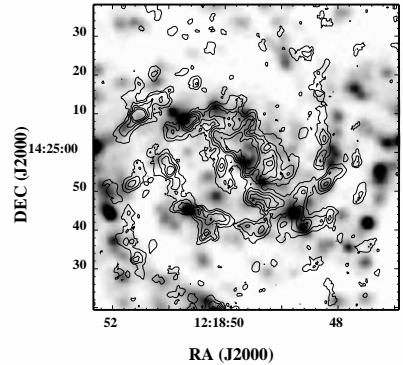}
  \end{center}
  \caption{Projected image of NGC 4254: 
CO contours on an H$\alpha$ image.}
\label{fig.COHa}
\end{figure}
 We show an overlaid image of CO and H$\alpha$ in 
Figure \ref{fig.COHa}.

 As we see in this figure, 
it is nearly face-on (We derived an inclination angle as 34$^\circ$ 
for central region. 
 See the following paragraph for detail.) 
and has well-ordered spiral arms 
(classified as arm class 9 by \citet{EE87}), 
so that the spiral structure can be easily traced. 
 This is why we selected this galaxy to 
examine offsets between H$\alpha$ and CO arms.

 Position angle (P.A.) and inclination ({\it i}) are 
also important parameters of a galaxy. 
 We determined them 
applying the task 'GAL' of AIPS to the velocity field of CO.
 This task cuts annuli out from a velocity field, 
and fits them with a pure circular rotation velocity field 
in order to derive P.A. and {\it i} at each radius.
 We averaged the derived values at $5'' < r < 16''$,
corresponding to 0.4 kpc $< r <$ 1.2 kpc, 
where the values do not change largely,
and then obtained 
P.A. = $72^\circ \pm 4^\circ$ and $i = 34^\circ \pm 5^\circ$. 
 In Table \ref{tb.pa_i}, these P.A. and {\it i} are 
listed with values from other papers.
\begin{table}
\begin{center}
\caption{Position angle and inclination of NGC 4254\label{tb.pa_i}}
\begin{tabular}{ccc}
\hline\hline
 & P.A. [$^{\circ}$] & {\it i} [$^{\circ}$]\\
\hline
   \citet{Schw76} & 58 & 29\\
   \citet{IOHW} & 70 & 42\\
   \citet{Phoo93} & 68 & 42\\
   This letter (2003) & 72 & 34\\
\hline
\end{tabular}
\end{center}
\end{table}
 Although only the central region was used in this work,
the determined values are not so different from previous work, 
in most of which the whole image of the galaxy was used for 
the determination. 

 Using the obtained P.A. and {\it i},
we calculated the rotation velocity of gas 
by the use of 'GAL' again. 
\begin{figure}
  \begin{center}
    \FigureFile(60mm,60mm){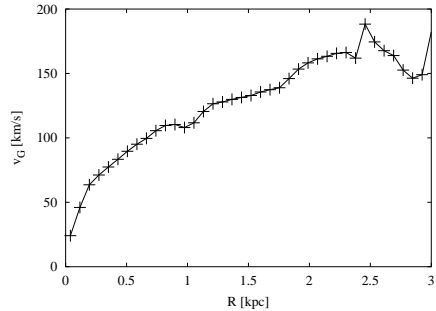}
  \end{center}
  \caption{Rotation curve of NGC 4254, determined by CO velocity field.}
\label{fig.RC}
\end{figure}
 The resultant rotation curve (Figure \ref{fig.RC}) 
is not perfectly the same with that derived 
by the iteration method \citep{Tak02, ViCS2}, 
since they used the different value of P.A. and {\it i}.

\begin{figure}
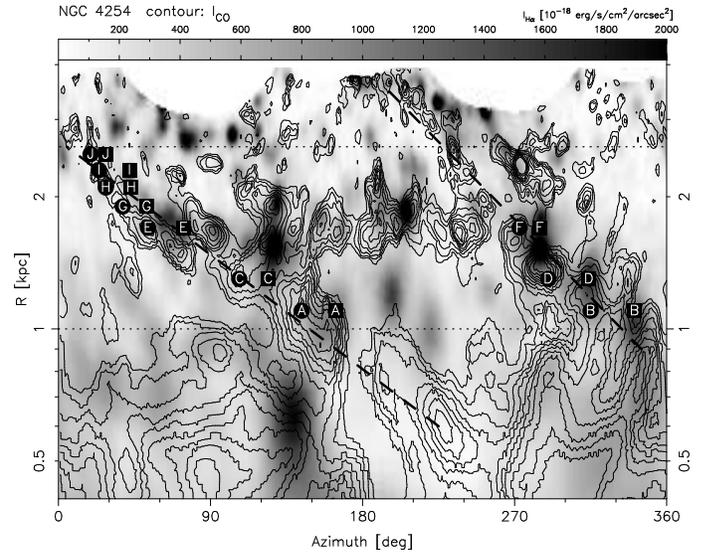

  \begin{center}
    \FigureFile(90mm,90mm){phase.ps}
  \end{center}
  \caption{Phase diagram of NGC 4254: CO contours on H$\alpha$ image.
 Thick dashed lines indicate rough positions of two marked spiral arms.
 Filled circles and boxes are the peaks of CO arm and H$\alpha$ arm, 
respectively, which are defined by the analysis described in section 4.
 Labels, from A to J, are to identify these peaks 
and offsets presented in Figure \ref{fig.chi_fit}.
 The region between the two horizontal lines (1 kpc $< r <$ 2.6 kpc) 
indicates where we used in the analysis.}
\label{fig.phase}
\end{figure}

 These P.A. and {\it i} were also used to deproject the image into 
a face-on view and to transform 
it into a polar-coordinate image, 
a phase diagram (Figure \ref{fig.phase}).
  Azimuth is taken to be zero at the NW side of minor axis 
and to increase clockwise, 
in the same direction of this galaxy's rotation 
assuming the trailing arms.
 A range between the two horizontal lines in this figure 
(1 kpc $< r <$ 2.6 kpc) show the region used in the following analysis. 
 Filled boxes and circles are the peaks of H$\alpha$ 
and CO intensity, respectively.
 See the following section for detail.

\section{Application and Result}
\subsection{Deriving Offsets}
 We divided the phase diagram into 
strips of 200 pc radial width, 
averaged the intensity with 
respect to radius at each azimuth in each strip, 
and plotted the averaged intensity against azimuth.
 We defined an offset angle $\theta$ as an azimuthal angular separation of 
intensity peaks between the H$\alpha$ and CO arms.
 Since arms of this galaxy are not totally continuous 
and have some breaks, 
we could not find corresponding peaks at some radii.
 In addition, 
we only used the
range of 1 kpc $< r <$ 2.6 kpc, 
since the spiral feature is not clear in the inner region 
and the rotation curve has a large dispersion in the outer region 
(See Figure \ref{fig.RC}).
 In Figure \ref{fig.phase}, this range is indicated by 
two horizontal lines. 
 The peak of CO and H$\alpha$ arm is marked 
by a filled and labeled 
circle and box, respectively, with a label from A to J, 
corresponding to that in Figure \ref{fig.chi_fit}.

 The largest factor for the error in $\theta$ is 
the resolution of CO data, which is about 2 arcsec.
 Thus this error is written as 
\begin{equation}
\Delta \theta  \simeq 8.9\times \Big(\frac{R}{\rm kpc}\Big)^{-1}
\quad {\rm [deg]}.
\label{eq.delta}
\end{equation}
 In addition, the uncertainty of the coordinate fitting of H$\alpha$ image 
and the fact that the determination of offset angles 
is somewhat subjective 
would also be possible sources of error in $\theta$.
 We, however, neglected these errors for the simplicity of 
calculations, so that 
the error in $\theta$ could be larger than 
the value obtained from equation (\ref{eq.delta}) 
by a factor of 2 at most.

\subsection{Fitting}
\begin{figure}
  \begin{center}
    \FigureFile(80mm,80mm){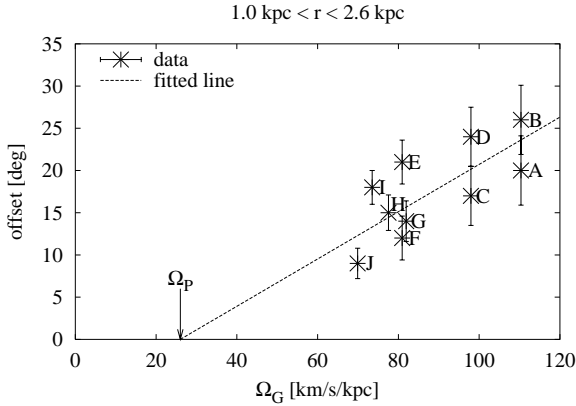}
  \end{center}
  \caption{The plot of offset angles against 
the angular rotation velocity.
 Labels, from A to J, are to identify the offsets, 
corresponding to the intensity peaks in Figure \ref{fig.phase}.
 Errorbars are calculated from equation (\ref{eq.delta}) 
and the dashed line is the fitted line obtained 
by the $\chi$-square method. 
 The horizontal-axis-intercept and the gradient of this line 
corresponds to $\Omega_{\rm P}$ and $t_{\rm H\alpha}$, respectively.}
\label{fig.chi_fit}
\end{figure}

 We plotted the derived offsets against $\Omega_{\rm G}$, 
derived from rotation curve (Figure \ref{fig.chi_fit}). 
 Errorbars in offset angle are $\Delta \theta$, calculated from 
equation (\ref{eq.delta}) at each radius. 
 We used the $\chi$-square fitting method and 
obtained $t_{\rm H\alpha} = ( 4.8 \pm 1.2 )\times 10^6$ yr 
and $\Omega_{\rm P} = 26^{+10}_{-6} ~{\rm km~s^{-1}~kpc^{-1}}$.
 The dashed line in Figure \ref{fig.chi_fit} is the fitted line, and 
the gradient and horizontal-axis-intercept of this line corresponds to 
the resultant value of $t_{\rm H\alpha}$ and $\Omega_{\rm P}$, 
respectively.
 
 \citet{Kranz01} showed that the corotation radius for the same 
galaxy is about 7.5 kpc from numerical simulations, 
and corresponding $\Omega_{\rm P}$ is about 20 
${\rm km~s^{-1}~kpc^{-1}}$.
 They assumed the distance to the galaxy as 20 Mpc.
 If we adjust this result to the distance of 16.1 Mpc, 
$\Omega_{\rm P}$ should be about 25 ${\rm km~s^{-1}~kpc^{-1}}$.
 The typical timescale of star formation has been 
suggested as about $10^7$ yr from the calculation of 
Jeans time in molecular clouds. 
 Our results are in good agreement with these 
previous studies.

\subsection{Discussion}
 In our method, we assumed a pure circular rotation 
and a constant time delay for star formation from 
molecular clouds,
and we did not do any correction for extinction in H$\alpha$ data.
 We however know that deviations from these assumptions 
are not critically large.
 The streaming motion and the velocity dispersion 
would generate some non-circular motion, 
but in a spiral disk, these are usually about 10 km s$^{-1}$ 
\citep{AW96, CB97}, 
which is very small compared to the circular rotation velocity of 
$\sim 100$ km s$^{-1}$.
 The validity of the latter assumption, constant $t_{\rm H\alpha}$, 
can be checked by the linearity of 
$\theta-\Omega_{\rm G}$ plot (Figure \ref{fig.chi_fit}). 
 Since this plot is well fitted with a line, 
dependence of $t_{\rm H\alpha}$ on the environment should be small 
and included in the error of the derived value.

 The extinction of H$\alpha$ is surely significant to luminosity 
and this might lead us to overestimate the offset values, 
since the amount of extinction becomes larger as H{\sc ii} regions 
get closer to molecular arms and an H$\alpha$ arm 
thus seems to shift to the downstream side from where it really is.
  \citet{GG96} showed azimuthal brightness profiles of NGC 4254 
in $K_s$-, $g$-, and $r_s$-band image at R=$75''$ 
(Figure 20 in their paper).
 Though the shape of profile is different from one to another, 
the positions of peak brightness are well consistent among them. 
 Hence, we can assume that the extinction would hardly change 
the position of H$\alpha$ arms 
and neglect the effect of extinction on the offset values.

\section{Conclusion}
 We examined the offset between the arm of H$\alpha$ and 
CO of SAc galaxy NGC 4254, derived ten offset values 
and found the linear relation between 
the offset angle and the angular rotation velocity.
 This linearity implies that 
a rigid spiral pattern exists and that 
the physical processes of star formation can be 
regarded as not to extremely change in the disk.
 We emphasize that this is the first work to show 
this relationship. 

 With this relation, we derived the pattern speed 
$\Omega_{\rm P} =26^{+10}_{-6} ~{\rm km~s^{-1}~kpc^{-1}}$, 
and the time delay for star formation 
$t_{\rm H\alpha} = ( 4.8 \pm 1.2 )\times 10^6$ yr.  
 We again emphasize that we can derive both $\Omega_{\rm P}$ 
and $t_{\rm H\alpha}$ simultaneously.
 Since the obtained values are consistent with previous studies,  
the way of our analysis can be a new method for the determination of 
the pattern speed 
and the typical timescale needed for star formations. 
 Thus, this method can investigate not only 
the kinematics, but also 
the star formation mechanisms quantitatively. 
 Moreover, this method can be applied to many spiral galaxies, 
since offsets between H$\alpha$ and CO arms have been found in 
many spirals.

\bigskip
 We are very grateful to Rebecca A. Koopmann, Jeffrey D. P. Kenney, 
and Judith Young for kindly providing us 
with their $R$ and H$\alpha$ image of NGC 4254.
 H. N. is financially supported by a Research Fellowship from the 
Japan Society for the Promotion of Science for Young Scientists.


\end{document}